\documentclass{aa} 
\usepackage{float} 
\usepackage{graphicx}  
\usepackage{txfonts}  
\usepackage{natbib}
\usepackage{float}
\bibpunct{(}{)}{;}{a}{}{,}
\usepackage[usenames,dvipsnames]{xcolor}
\usepackage[normalem]{ulem}

\begin{document}  

 
\title{Impacts of radiative accelerations on solar-like oscillating main-sequence stars}
  
  \author{ M. Deal\inst{1}, G. Alecian\inst{2}, Y. Lebreton\inst{1,3}, M. J. Goupil\inst{1}, J. P. Marques\inst{4}, F. LeBlanc\inst{5}, P. Morel\inst{6} \and B. Pichon\inst{6}}
\institute{LESIA, Observatoire de Paris, PSL Research University, CNRS, Universit\'e Pierre et Marie Curie, Universit\'e Paris Diderot,
92195 Meudon, France\
          \and
           LUTH, Observatoire de Paris, PSL Research University, CNRS, Universit\'e Pierre et Marie Curie, Universit\'e Paris Diderot,
92195 Meudon, France\
			\and
           Institut de Physique de Rennes, Universit\'e de Rennes 1, CNRS UMR 6251, 35042 Rennes, France\	 
           \and
		Institut d’Astrophysique Spatiale, UMR8617, CNRS, Universit\'e Paris XI, B\^atiment 121, 91405 Orsay Cedex, France\
		   \and
		D\'epartement de physique et d'astronomie, Universit\'e de Moncton, Moncton, N.-B., E1A 3E9, Canada\ 
		\and
		Universit\'e de Nice Sophia-Antipolis, OCA, Laboratoire
Lagrange CNRS, BP. 4229, 06304, Nice Cedex, France\   	
            \email{morgan.deal@obspm.fr} 
           }
           
\date{\today}

\abstract
{Chemical element transport processes are among the crucial physical processes needed for precise stellar modelling.  Atomic diffusion by gravitational settling nowadays is usually taken into account, and is essential for helioseismic studies. On the other hand, radiative accelerations are rarely accounted for, act differently on the various chemical elements, and can strongly counteract gravity in some stellar mass domains. The resulting variations of the abundance profiles may significantly affect the structure of the star.}
{In this study we aim at determining whether radiative accelerations impact the structure of solar-like oscillating main-sequence stars observed by asteroseismic space missions.}
{We implemented the calculation of radiative accelerations operating on C, N, O, Ne, Na, Mg, Al, Si, S, Ca, and Fe in the CESTAM code using the Single-Valued Parameter method. We built and compared several grids of stellar models including gravitational settling, but some with and others without radiative accelerations. We considered masses in the range $[0.9, 1.5]\ M_\odot$ and 3 values of the metallicity around the solar one. For each metallicity, we determined the range of mass where differences between models due to radiative accelerations exceed the uncertainties of global seismic parameters of the \textit{Kepler Legacy} sample or expected for PLATO observations.}
{We found that radiative accelerations may not be neglected for stellar masses larger than 1.1~M$_{\odot}$ at solar metallicity. The difference in age due to their inclusion in models can reach 9\% for the more massive stars of our grids. We estimated that the percentage of the PLATO core program stars whose modelling would require radiative accelerations ranges between 33 and 58\% depending on the precision of the seismic data.}
{We conclude that, in the context of Kepler, TESS, and PLATO missions, which provide (or will provide) high quality seismic data, radiative accelerations can have a significant effect when inferring the properties of solar-like oscillators properly. This is particularly important for age inferences. However, the net effect for each individual star results from the competition between atomic diffusion including radiative accelerations and other internal transport processes. Rotationally induced transport processes for instance are believed to reduce the effects of atomic diffusion. This will be investigated in a forthcoming companion paper.}

\keywords{asteroseismology -- diffusion -- stars: interiors -- stars: evolution}
  
\titlerunning{Impacts of radiative accelerations on solar-like oscillating main-sequence stars}
  
\authorrunning{Deal et al.}  

\maketitle 

\section{Introduction} 

Understanding and modelling the transport of chemical elements inside stars still remain a difficult challenge for the theory of stellar structure and evolution. Chemical abundances indeed play an important role for  determining  the structure and evolution of stars. The internal distribution of chemical elements results from the competition of several transport processes within the star which are still barely understood and/or poorly modelled. 

Transport processes can be constrained using photospheric observations, but the impact on the internal structure can only be probed using stellar oscillations. The CoRoT \citep{baglin13} and \textit{Kepler} \citep{gilliland10} space missions provided a wealth of high quality photometric light curves. Seismic data derived from these observations improved the characterisation of the observed main sequence stars and provide constraints on their internal structures \citep[for reviews, see][]{chaplin13, deheuvels16, christensen16}.
 
The PLATO ESA mission \citep{rauer14} will be launched in 2026 and offers a new perspective to constrain further our stellar evolution models. The objectives of the project are the detection, the full characterization of Earth-like planets orbiting solar-like stars and the study of the evolution of star-planet systems. While the detection of exoplanets requires very high signal-to-noise ratios and long observing time, the full characterization of these detected objects requires the precise determination of the stellar parameters of the host-stars. The PLATO mission aims at observing a large number of stars while combining two techniques:
\begin{itemize}
\item the detection by photometric transit and a ground based follow up in radial velocity which will provide the planet-to-host star radius and mass ratios respectively;
 
\item asteroseismology analysis(coupled with spectroscopic observations) which will provide precise masses, radii and more importantly ages of the host stars.
\end{itemize}

The goal is to reach uncertainties of the order or less than 3\% in radius, and 10\% in mass for the planets. This translates into the need to reach uncertainties of the order of or less than 2\% in radius, and 15\% in mass for the host-stars. A PLATO objective is also to reach an uncertainty as small as 10\% for the age determination of a solar-like host-star. The current stellar models are still not able to provide such accuracy.

The study of the competition between microscopic and macroscopic transport processes is a necessary step towards more accurate stellar models. Helioseismology showed the necessity to include atomic diffusion to properly model the Sun \citep{christensen93}. It is a microscopic process which occurs in every star due to the gradients of $T$, $P$, ... etc. This process was first discussed by \cite{eddington26} and the importance of radiative accelerations was first recognized by \cite{michaud70} and \cite{watson71}. The diffusion velocity of an element mainly depends on two forces (or accelerations) : gravity, which makes the element migrate toward the center of the star, and radiative accelerations which generally push the element up toward the surface. This latter is due to the capability of ions to absorb photons (according to their atomic properties) and to acquire part of their momentum. Atomic diffusion principally results from the competition between these two forces. 

For G, F, and late A type main-sequence stars (Population I and II), models including atomic diffusion may produce too large depletions or accumulations of chemical elements if no additional mixing other than convection is considered. This is the reason why models need to include additional macroscopic transport processes to reproduce the observed surface abundances \citep[e.g.][]{korn07}. Atomic diffusion can then be used as a proxy to determine the efficiency of macroscopic transport processes or the rate of mass loss needed to reproduce observations and then predict which processes play a role \citep[e.g.][]{talon06,michaud04,michaud11}. 

Atomic diffusion leads to local modifications of the abundance profiles and, hence, to a modification of the Rosseland opacities. This has important structural effects in stars as for example the opacity-induced iron/nickel convection zone triggered by the local accumulation of these species around $200~000~K$ and where these elements are main contributors to the opacity in F and A type stars \citep{richard01,theado09,deal16}. This opacity modification close to the bottom of the surface convection zone also causes an increase of the mass of the surface convection zone in F type stars \citep{turcotte98b}. The local accumulation of elements may also lead to an inverse mean molecular weight gradient which triggers thermohaline (or fingering) convection in F and A type stars \citep{theado09,deal16} and in B type stars \citep{hui-Bon-Hoa18}. It was shown that neglecting radiative accelerations in the modelling of 94 Ceti A (a F-type star showing solar-like oscillations) using asteroseismic data leads to a 4\% age difference \citep{deal17}. 

Nowadays only a few evolution codes incorporate consistent computations of stellar models including the complete treatment of atomic diffusion. The Montreal/Montpellier code \citep{turcotte98} computes radiative accelerations using OPAL monochromatic data and the opacity sampling method \citep[e.g.][]{leblanc00}. The Toulouse Geneva Evolution Code \citep{hui-bon-hoa08,theado12} includes the OPCD package\footnote{http://cdsweb.u-strasbg.fr/topbase/testop/TheOP.html} from the Opacity Project calculations \citep{seaton05} for the opacities and computes radiative accelerations using the Single-Valued Parameter (SVP) approximation proposed by \cite{alecian02} and \cite{leblanc04}. The SVP approximation allows very fast computations with no need for monochromatic data as they are tabulated within the method. The MESA code computes Rosseland mean opacities and radiative accelerations with the OPCD3 method \citep{paxton18}, optimised by the work of \cite{hu11}. In the present paper, we add to the above list the CESTAM code \citep{marques13} where we implemented the radiative accelerations within the framework of the SVP approximation while using the OPCD3 package for calculations of opacities.

Atomic diffusion has an important impact on the structure of stars. The effects are detectable in the Sun. It has also been shown to play a role in several other type of pulsating stars \citep{charpinet97,turcotte00,alecian09,theado12bisbis}. We aim here at determining whether atomic diffusion, including the effect of radiative accelerations, needs to be taken into account in the modelling of solar-like oscillating main sequence stars. This is a prerequisite for an optimal interpretation of the data provided by CoRoT and \textit{Kepler} and by future space mission such as TESS and PLATO. Macroscopic transport processes such as those induced by turbulent convection and/or rotation also play an important role, and the competition with atomic diffusion is not straightforward; several parameters come into play and the net result likely depends on the type of stars, if not on the specificities of each individual star. We have therefore started an in-depth study which ultimately ought to provide the net result of this competition on the transport of chemical elements and the associated consequences on the structure, evolution of the star and its solar-like oscillating properties. The present paper represents the first step of this study. Our purpose here is a theoretical quantification of the sole impact of atomic diffusion - more specifically the radiative acceleration process - on the structure, surface abundances and some basic seismic properties of stars. No  macroscopic processes other than convection is taken into account. The results presented here may then be interpreted as the maximum impact of atomic diffusion including radiative accelerations. The inclusion of the competitive effect of rotationally induced mixing as allowed by our evolutionary code is in progress and will constitute the second paper of the series.

The paper is organized as follows: we first detail the new developments of the CESTAM code in Section 2. Section 3 then presents the grids of stellar models which focus on low mass main-sequence stars and the impact of the radiative accelerations on the stellar structure and chemical abundances by comparing models computed with and without radiative accelerations. Some seismic implications are presented in Section 4. The impact of the radiative acceleration on the surface iron abundance and thereby on the stellar characterisation are discussed in Section 5, while Sections 6 and 7 are devoted to discussions and conclusions respectively.
\section{CESTAM stellar models}

\subsection{Standard physics}
\label{sec2.1}
The stellar models are computed using the CESTAM code \citep{marques13} which is based on the CESAM code \citep{morel97,morel08}, with a more detailed treatment of rotationally induced transport processes. Here we don't consider the effect of rotation. A second forthcoming paper will discuss the net results of the competition between atomic diffusion (including radiative accelerations) and rotationally induced transport of angular momentum and chemical elements.  

The CESTAM models can be computed using the opacities given by the OP \citep{seaton05} or OPAL \citep{iglesias96} tables complemented at low temperature by the Wichita opacity data \citep{ferguson05}. The equation of state is the OPAL2005 one \citep{rogers02}. The nuclear reactions are taken from the NACRE compilation \citep{angulo99} except for the $^{14}$N(p,$\gamma$)$^{15}$O reaction, for which we used the LUNA reaction rate given in \cite{imbriani04}. The convection was treated following \citet[][hereafter CGM]{canuto96} with a mixing-length $l=\alpha_\mathrm{CGM} H_P$, where $H_P$ is the pressure scale height. We took into account the overshooting of the convective core, with an overshoot extent of $0.15 \times \min(H_p, r_{cc})$ where  $r_{cc}$ is the radius of the Schwarzschild's convective core. This choice is compatible with recent determinations of the overshooting extent based on the study of eclipsing binaries (Claret \& Torres, 2016) and on asteroseismology of solar-type stars (Deheuvels et al, 2015). The atmosphere is computed in the gray approximation and integrated up to an optical depth of $\tau = 10 ^{-4}$ and no mass loss was taken into account. We used the solar mixture of \cite{asplund09} with meteoritic abundances for refractory elements as recommended by \cite{serenelli10}.

In CESTAM two formulations are available for atomic diffusion: the first one is based on the work of \cite{michaud93} (hereafter MP93) and the second one on the Burgers equations \citep{burgers69}. Here we used the  \cite{michaud93}'s formulation. The MP93 approximation used in the CESTAM code considers the diffusion of trace elements (with partial ionisation) in a fully ionised plasma of H and He. This is an approximation of the Burgers' equations. Some comparisons were made with the full Burgers' treatment for the Sun \citep{turcotte98}, and in the framework of the Evolution and Seismic Tool Activity (ESTA) for the CoRoT mission for the effect of gravitational settling only \citep{thoul07, montalban07, lebreton08}. The advantage of the MP93's method is that computational times are very short.

\subsection{Partial ionisation}
\label{partialioni}

Partial ionisation, which is often not considered in evolution codes, is extremely important for atomic diffusion calculations \citep{montmerle76,michaud15}. Firstly, this is because radiative acceleration depends on atomic properties of ions, and secondly because the diffusion velocity is proportional to the diffusion coefficient ($D_{ip}$), which is proportional to ${Z_i}^{-2}$ (where ${Z_i}$ is the electric charge of the ion in proton charge units). Hence, for instance, for two ions with respective charges  ${Z_i}$ of 5 and 6 undergoing the same resultant acceleration, in the same stellar layer, the velocity of the ion with charge 6 is 30\% smaller than that of the ion with charge 5. Another example: assuming that iron is fully ionized in diffusion velocity calculations around the depth where the iron opacity bump occurs ($\log{T}\approx 5.2$) gives erroneous velocity estimation by more than a factor of 10! The error made by assuming full ionisation in atomic diffusion velocity calculations is larger for stars with a small surface convection zone (larger $T_{\rm{eff}}$) since ions have smaller ${Z_i}$ at its bottom (cooler layers). Therefore, neglecting partial ionisation in diffusion calculations of chemical elements leads to large underestimates for the diffusion velocities. In this study, partial ionisation on heavy elements is taken into account through an average electric charge $\bar{Z}_i$ (instead of ${Z_i}$) for each element. This simplifies significantly the numerical treatment of the diffusion equations (see Section~\ref{diffeq}), since one does not need to consider individual ions (the same approximation is used in \citealt{turcotte98}). Hereafter, $i$ represents an element whose atoms locally possess an average electric charge $\bar{Z}_i$ depending on the local plasma conditions.

\subsection{Diffusion equation}\label{diffeq}

The equation describing the evolution of the chemical composition reads:
\begin{equation}
\label{equadiff}
\rho\frac{\partial c_i}{\partial t}=-\frac{1}{r^2}\frac{\partial}{\partial r}\left[r^2\rho D_{turb}\frac{\partial c_i}{\partial r}\right]-\frac{1}{r^2}\frac{\partial}{\partial r}[r^2\rho v_{i} c_i] -\lambda_i c_i
\end{equation}
\noindent where $c_i$ is the concentration of element $i$, $\rho$ is the density in the considered layer, $D_{turb}$ is a turbulent diffusion coefficient, and $\lambda_i$ is the nuclear reaction rate related to the element $i$. 
In Eq.\ref{equadiff}, $v_i$ is the atomic diffusion velocity that can be expressed in the case of a trace element $i$ as 
\begin{equation}
v_i=D_{ip}\left[-\frac{\partial \ln c_i}{\partial r}+\frac{A_i m_p}{k T}(g_{rad,i}-g)+\frac{({\bar{Z}_i}+1)m_p g}{2 k T}+\kappa_T\frac{\partial \ln T}{\partial r}\right]
\label{eqvdiff}
\end{equation}

\noindent where $D_{ip}$ is the diffusion coefficient of element $i$ relative to protons, and $A_i$ is its atomic mass. The variable $g_{rad,i}$ is the radiative acceleration on element $i$, $g$ is the local gravity, $\bar{Z}_i$ is the average charge (in proton charge units) of element $i$ (it is roughly equal to the charge of the "dominant ion"), $m_p$ is the mass of a proton, $k$ is the Boltzmann constant, $T$ is the temperature and $\kappa_T$ is the thermal diffusivity. It should be noted that $\bar{Z}_i$ is used when estimating $D_{ip}$.

The competition between macroscopic transport processes and atomic diffusion is given by the first two right-hand side terms of Eq.~\ref{equadiff}.  

\subsection{Radiative accelerations in CESTAM}\label{newdev}
\begin{figure*}[h!]
\center
\includegraphics[scale=0.5]{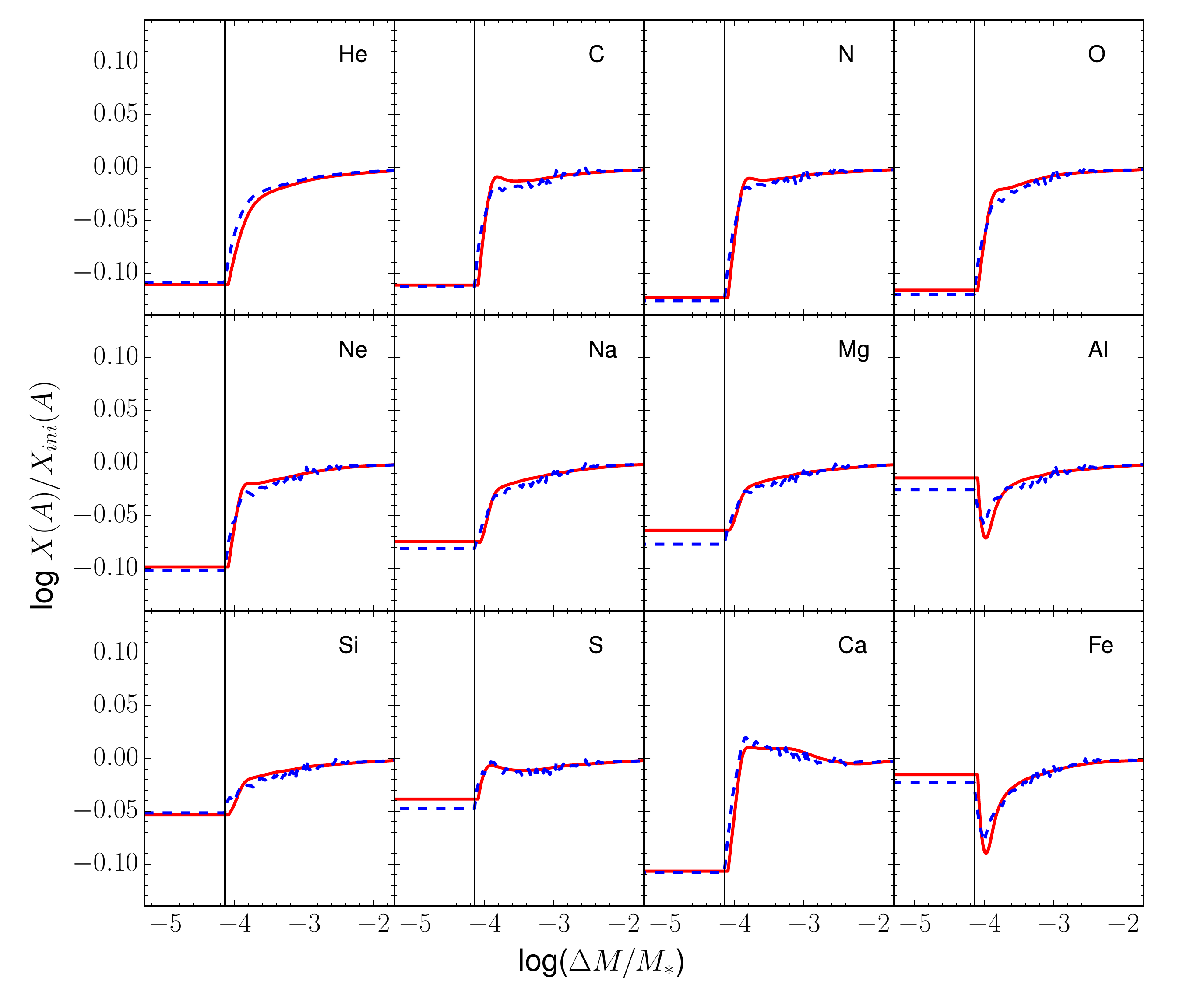}
\caption{Comparisons of abundance profiles for different elements according to $\log (\Delta M/M_*)$ (where $\Delta M$ is the mass between the considered layer and the surface) for a 1.4~M$_{\odot}$ at Z=0.025 and 400 Myr between a model computed with the Montreal/Montpellier code (blue dashed curves) and the CESTAM code (red solid curves). The solid vertical lines indicate the position of the bottom of the surface convection zone of the models. }
\label{fig:1}
\end{figure*}

\subsubsection{Atomic diffusion}
In some evolution codes including atomic diffusion, one considers a mixture of hydrogen, helium, and of a mean heavy element (with mass fractions X, Y, Z respectively) (e.g. \citealt{thoul94}). This (X, Y, Z) mixture treatment of atomic diffusion gives acceptable results (depending on the needed accuracy) for stars with masses close to that of the Sun, i.e. in stars where radiative accelerations are systematically weak compared to gravity (i.e. gravitational settling is dominant). But this approximation is no longer valid for more massive stars where radiative accelerations dominate gravity. In this case, the migration of chemical elements is often towards the surface, depending on the interaction of their ions with the radiation flux. The sign and intensity of the diffusion velocity of a given species depends on the atomic properties of the dominating ions, and on depth (or local physical conditions). This is why elements cannot be treated as a unique mean heavy element Z.

In its present version, CESTAM computes the evolution of  abundances of all the elements available in the OPCD3 package \citep{seaton05} and of some isotopes: H, $^3$He, $^4$He, $^{12}$C, $^{13}$C, $^{14}$N, $^{15}$N, $^{15}$O, $^{16}$O, $^{17}$O, $^{22}$Ne, $^{23}$Na, $^{24}$Mg, $^{27}$Al, $^{28}$Si, $^{31}$P (without radiative accelerations), $^{32}$S, $^{40}$Ca and $^{56}$Fe. It also takes into account the partial ionisation in computing diffusion velocities (see Eq.~\ref{eqvdiff}) which is a major new development in the evolution code under consideration. It is shown in the next sections that modifications of the structure and surface abundances of stars occur when $\bar{Z}_i$ is used instead of the charge of the fully ionised element.

Radiative accelerations in CESTAM are computed using the Single-Valued Parameter (SVP) approximations proposed by \cite{alecian02} and \cite{leblanc04}. This method is among the three ones generally used for radiative accelerations (see Alecian 2018): (i) direct use of atomic data (the most accurate method, but the most heavy one to carry out), (ii) use of opacity tables with fixed frequency grid (less accurate, but numerically lighter), (iii) use of parametric approximations (less accurate than (ii), but extremely fast, numerically). The first method is generally used to compute radiative accelerations in stellar atmospheres \citep{hui-bon-hoa00, alecian04a, alecian06, leblanc09} and necessitates direct integration over atomic transition profiles. The second one is valid for stellar interiors and is used in the Montreal/Montpellier code, and is also employed (with interpolation techniques) in the OPCD3 package \citep{seaton97, seaton07}. The third one corresponds to the SVP approximations and is only valid for stellar interiors.

The SVP method is based on a simplified form of the equations for radiative accelerations. They are obtained by separating the terms involving the atomic quantities from those describing the local plasma. SVP method needs very small tables, contrarily to the other methods. These small tables providing only six parameters per ion are pre-calculated for various stellar masses, and the numerical routines have to interpolate these data to fit the mass of the considered star (some tables may be found on the website http://gradsvp.obspm.fr, and a larger set of tables is in preparation). This method is numerically efficient and is tailored for use in stellar evolution codes.

The SVP method was implemented in the TGEC code \citep{theado12}, and we proceeded in the same way for its implementation in CESTAM using the same set of tabulated parameters as for TGEC. In this study, radiative accelerations are computed for C, N, O, Ne, Na, Mg, Al, Si, S, Ca and Fe. The SVP parameters have been calculated with the use of the Opacity Project data \citep{seaton92, cunto93}.

In order to avoid numerical instabilities due to sharp gradients of abundance produced by radiative accelerations, we added an ad-hoc turbulent mixing coefficient as done by \cite{theado09} and \cite{deal16}. The turbulent coefficient has the following expression:

\begin{equation}
\label{eq:1}
D_{turb}=D_{bcz,1}~\rm{exp} \left(\frac{r-r_{bzc}}{\Delta_1} \ln2\right)+D_{bcz,2}~\rm{exp} \left(\frac{r-r_{bzc}}{\Delta_2} \ln2\right)
,\end{equation}

where $D_{bcz}$ and $r_{bcz}$ are respectively the value of $D_{mix}$ and the value of the radius at the bottom of the convection zone. For the grids we choose $D_{bcz,1}=500$ $\rm{cm}^2~\rm{s}^{-1}$ and $\Delta_1=0.02$ of the radius of the star and $D_{bcz,2}=200$ $\rm{cm}^2~\rm{s}^{-1}$ and $\Delta_2=0.1$. This turbulent mixing coefficient was chosen in order not to affect significantly the evolution of the star and has a negligible effect on the results presented below. 

\subsubsection{Opacity tables}
In our models, atomic diffusion notably modifies the initial mixture of heavy elements in outer layers, which implies that pre-computed Rosseland opacity tables cannot be used throughout the interior and all along the evolution. We therefore had to recompute the Rosseland mean opacity locally at each timestep in the layers where the mixture changes much. For this purpose, we implemented in CESTAM the dedicated routine (mx.f) which handles the monochromatic opacity tables from the OPCD3 package \citep{seaton05}.

Since running the mx.f routine is time-consuming, we recomputed the Rosseland opacity only in the outer layers, when $\log(T)<\approx 6.23$. We point out that :
\begin{itemize}
\item at higher temperatures, to spare computing time, we used the pre-computed Rosseland mean OP opacity tables described in Sect.\ \ref{sec2.1}. 
\item at low temperatures ($T < 10^4$ K), the OPCD3 opacities are still available. Therefore, for consistency, we preferred to use them, rather than the more complete Wichita tables (which provide Rosseland means including molecular lines for a given mixture but are not available in the form of monochomatic opacities).
The impact of using OPCD instead of Wichita opacities in the low-temperature domain is that we miss the molecular contribution to the opacity. This may have some impact on the stellar properties especially for the colder stars. However, for these stars radiative accelerations are negligible and since our goal is to perform a relative comparison, this should not significantly modify our conclusions.
\end{itemize}

\subsection{Comparison and validation of the implementations}
To verify the validity of the new developments presented in Section~\ref{newdev}, we compared the results obtained with our new version of CESTAM to those obtained with the Montreal/Montpellier code. We chose a model of 1.4~M$_{\odot}$ with parameters listed in Table \ref{table:1}.

\begin{table}
\centering
\caption{Initial parameters of the comparison model} 
\label{table:1}
\begin{tabular}{l c c}
\hline
Model & CESTAM & Montreal/Montpellier \\
\hline
Mass (M$_{\odot}$) & 1.4  & 1.4\\ 
X$_{ini}$ & 0.69500 & 0.69500\\ 
Y$_{ini}$ & 0.27995 & 0.27995\\
Z/X$_{ini}$ & 0.0360 & 0.0360\\
$\alpha_{MLT}$ & $1.687$ & $1.687$ \\ 
Mixture & GN93 & GN93 \\ 
Opacities & OPCD+OPAL  & OPAL Mono \\
EoS & OPAL2005 & CEFF \\
Nuclear reactions & NACRE & Bahcall92\tablefootmark{a}\\ 
Core overshoot & none & none\\    
\hline
\end{tabular}
\tablefoot{\tablefoottext{a}{\cite{bahcall92}}}
\end{table}

Since the input physics of the models is not exactly the same (especially the equation of state and opacity tables) the structures are slightly different. Nevertheless, they are close enough for our purpose. 

Figure \ref{fig:1} shows the abundance profiles of various elements. The agreement between both codes is very satisfactory. The differences between them never exceed 3\% for the surface abundances. Elements are depleted/accumulated in the same way. We have also compared models for more massive stars, and the agreement is at the same level. Therefore, this comparison allows us to be confident in the use of this new version of the CESTAM code.
  
\section{Effects of atomic diffusion on the internal structure}\label{PLATO}
 
Our goal here is to evaluate the range of stellar mass and initial chemical composition for which radiative accelerations (hereafter g$_\mathrm{rad}$) cannot be neglected when computing accurately the structure and evolution of solar-like oscillating main-sequence stars. This will allow us to determine masses above which g$_\mathrm{rad}$ have to be taken into account to properly infer stellar parameters (age, mass, radius) from models. These are lower-limit masses because macroscopic transport processes apart from convection are not taken into account. This will allow to spare computational time when their effects are negligible.  For that purpose we built two sets of stellar model grids described below.
 
\subsection{Our grids of models}

\begin{figure*}
\center
\includegraphics[scale=0.45]{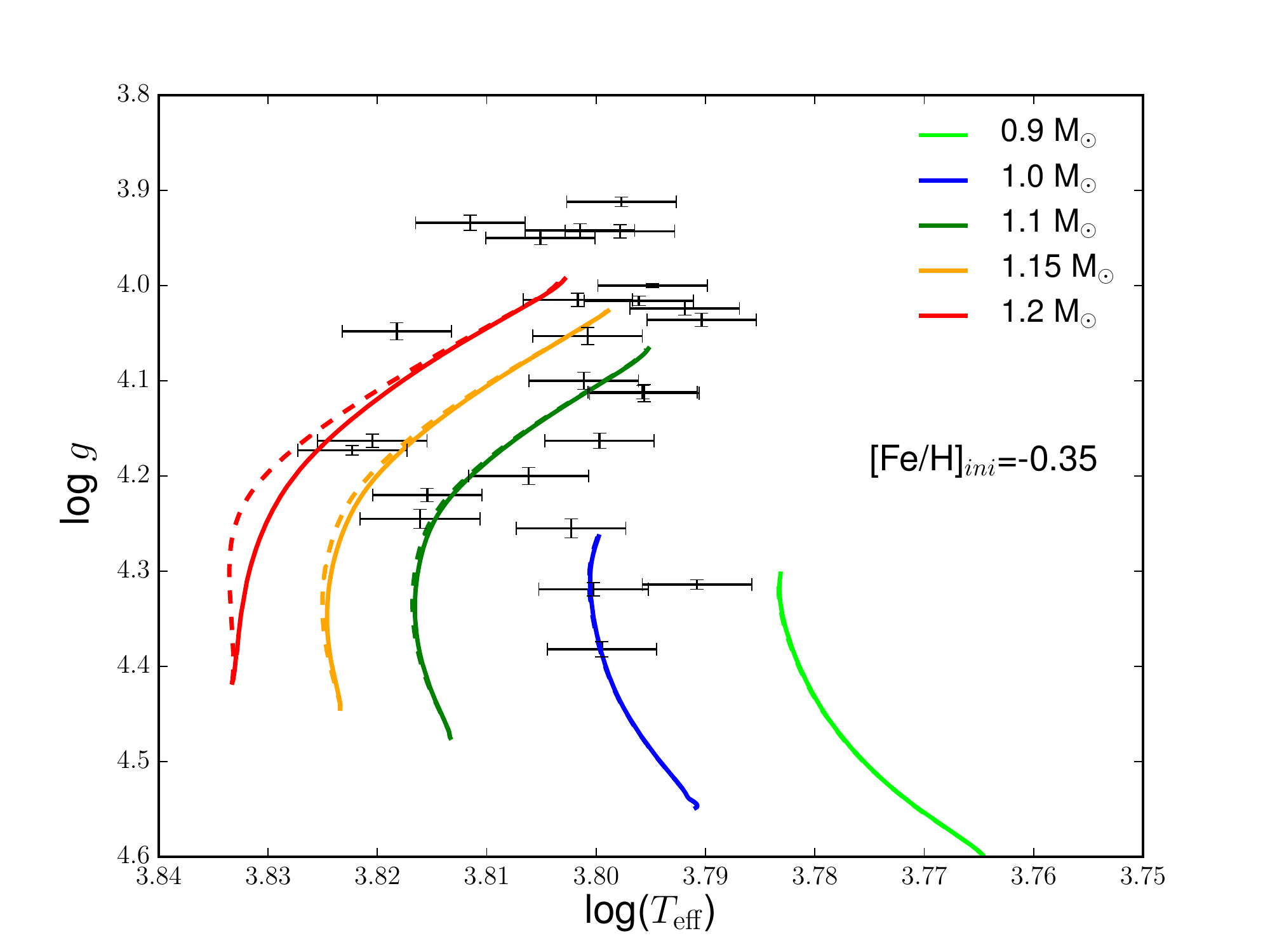}
\includegraphics[scale=0.45]{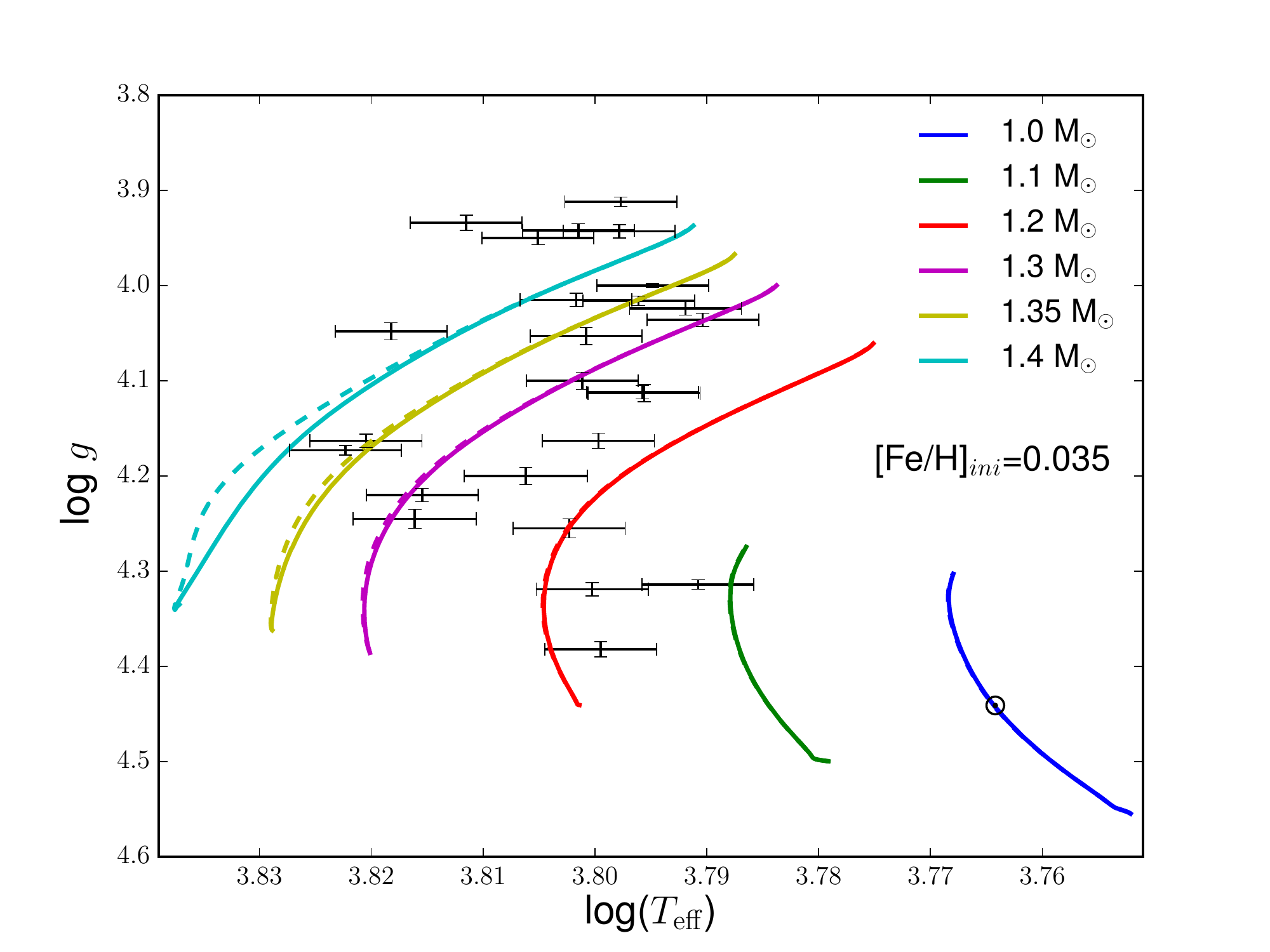}
\includegraphics[scale=0.45]{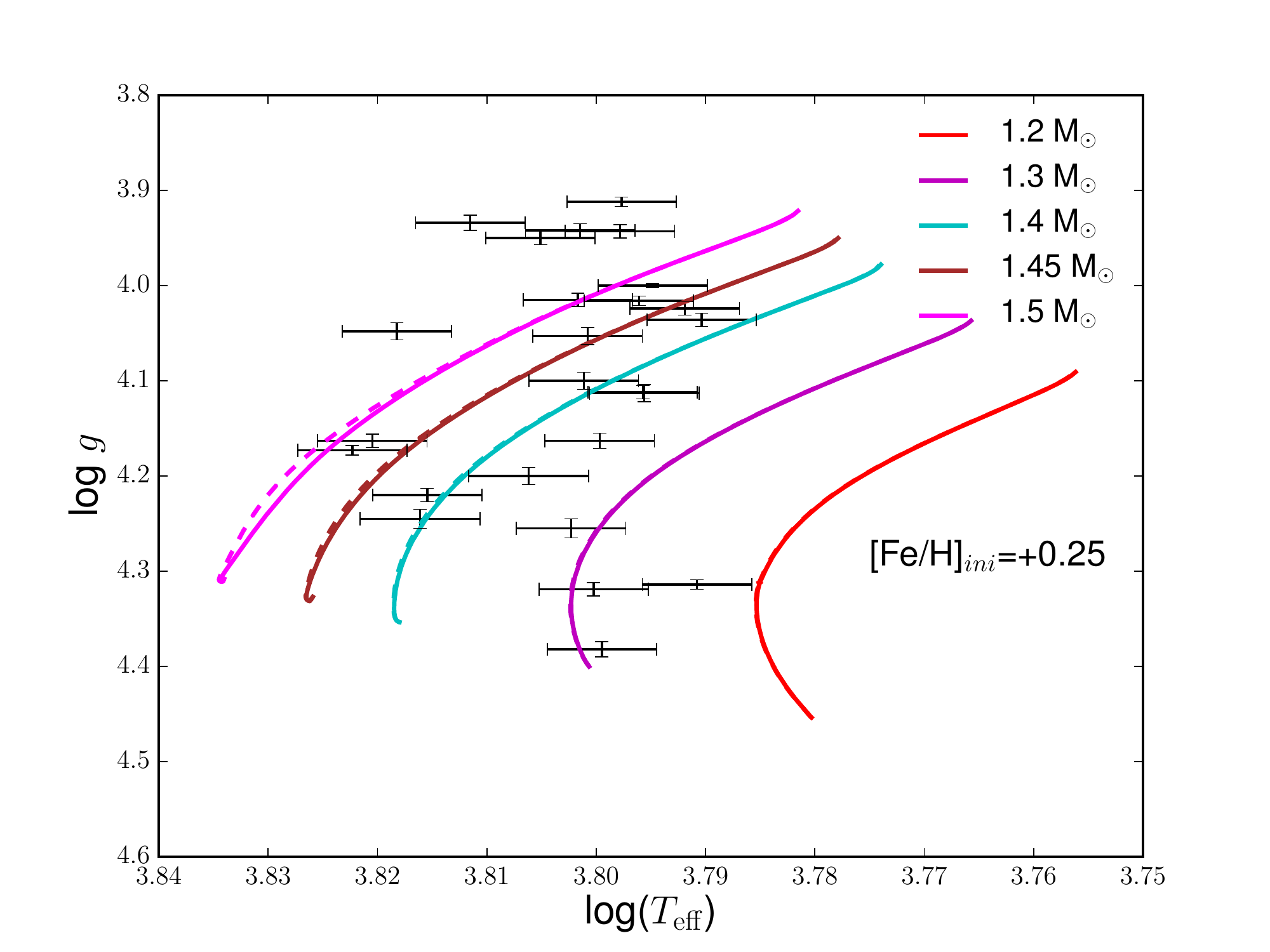}
\caption{HR diagrams for initial [Fe/H]$_{ini}$=-0.35 (upper left panel), [Fe/H]$_{ini}$=0.035 (upper right panel) and [Fe/H]$_{ini}$=+0.25 (lower panel). The dashed curves represent models without g$_\mathrm{rad}$ and the solid curves represent models including g$_\mathrm{rad}$. Black symbols are stars from the \textit{Kepler Legacy} sample \citep{lund17,silva17}.}
\label{fig:2}
\end{figure*}

We first define three grids of models listed in Table \ref{table:2}, each of them corresponding to a different metallicity. We have chosen masses in the range $[0.9, 1.5]\ M_\odot$, a range for which g$_\mathrm{rad}$ are expected to have the most significant impact on the structure and evolution of solar-like oscillating main-sequence stars.
In order to cover the wide range of metallicities of the CoRoT, Kepler, and in the future,  TESS and PLATO targets, we have considered three values of the initial metallicity for grids 1 to 3, respectively: $[\mathrm{Fe/H}]_{ini}= -0.35, +0.035$, and $+0.25$ dex, with: 
\begin{equation}\label{feh}
[\mathrm{Fe/H}]= \log(X_{Fe}/X_H)-\log(X_{Fe}/X_H)_\odot\,,
\end{equation}
\noindent where $X_H$ and $X_{Fe}$ are the hydrogen and iron abundances in mass fraction.
Models cover the whole main sequence life time, up to the stage where the central hydrogen content is $X_c=0.05$.

For each of these three grids, we have computed a first set of models including g$_\mathrm{rad}$, and a second set without g$_\mathrm{rad}$ (gravitational settling only) including only convection as macroscopic transport process.

The values of the mixing-length parameter $\alpha_\mathrm{CGM}$ and initial helium abundance $Y_\mathrm{ini}$ at solar metallicity were inferred from a solar model calibration. As g$_\mathrm{rad}$ are negligible in the Sun the calibration was done with gravitational settling only. A solar calibration consists in adjusting the initial helium abundance $Y_\mathrm{ini, \odot}$, metallicity $(Z/X)_\mathrm{ini, \odot}$, and $\alpha_\mathrm{CGM}$ of a $1\ M_\odot$ model so that it reaches at solar age, the observed solar luminosity, radius, and photospheric metallicity (see \citealt{morel08}). We obtained $Y_\mathrm{ini, \odot}=0.2578$ and $\alpha_\mathrm{CGM}=0.68$. From $Y_{ini, \odot}$ and a primordial helium abundance $Y_\mathrm{BB}=0.247$ \citep{peimbert07}, we obtained a helium to metal enrichment ratio $\Delta Y/\Delta Z= (Y_\mathrm{ini, \odot}-Y_{BB})/Z_\mathrm{ini, \odot}=0.9$ which we used to get the initial helium abundance for models with other metallicities. 

\begin{table}
\centering
\caption{Characteristics of the grids of models}
\label{table:2}
\begin{tabular}{l c c c }
\hline
Grid & 1 & 2 & 3 \\
\hline
[Fe/H]$_{ini}$ & -0.35 & 0.035 & +0.25 \\
Mass (M$_{\odot}$) & 0.9-1.2 & 1.0-1.4 & 1.0-1.5 \\ 
Step (M$_{\odot}$) & 0.05-0.1 & 0.05-0.1  & 0.05-0.1  \\ 
X$_{ini}$ & 0.7438 & 0.7280 & 0.7117\\ 
Y$_{ini}$  & 0.2503 & 0.2578 & 0.2655\\
(Z/X)$_{ini}$  & 0.0080 & 0.0195 & 0.0320 \\
$\alpha_\mathrm{CGM}$ & 0.68 & 0.68 & 0.68 \\   
\hline
\end{tabular}
\end{table} 

\subsection{Evolutionary tracks}\label{impact}
To characterize the differences between models with and without g$_\mathrm{rad}$ in the abundances and the structure of stellar interiors, we computed evolutionary tracks presented in Fig. \ref{fig:2} for the three grids of models described in Table~\ref{table:2}. 

\begin{figure*}
\center
\includegraphics[scale=0.5]{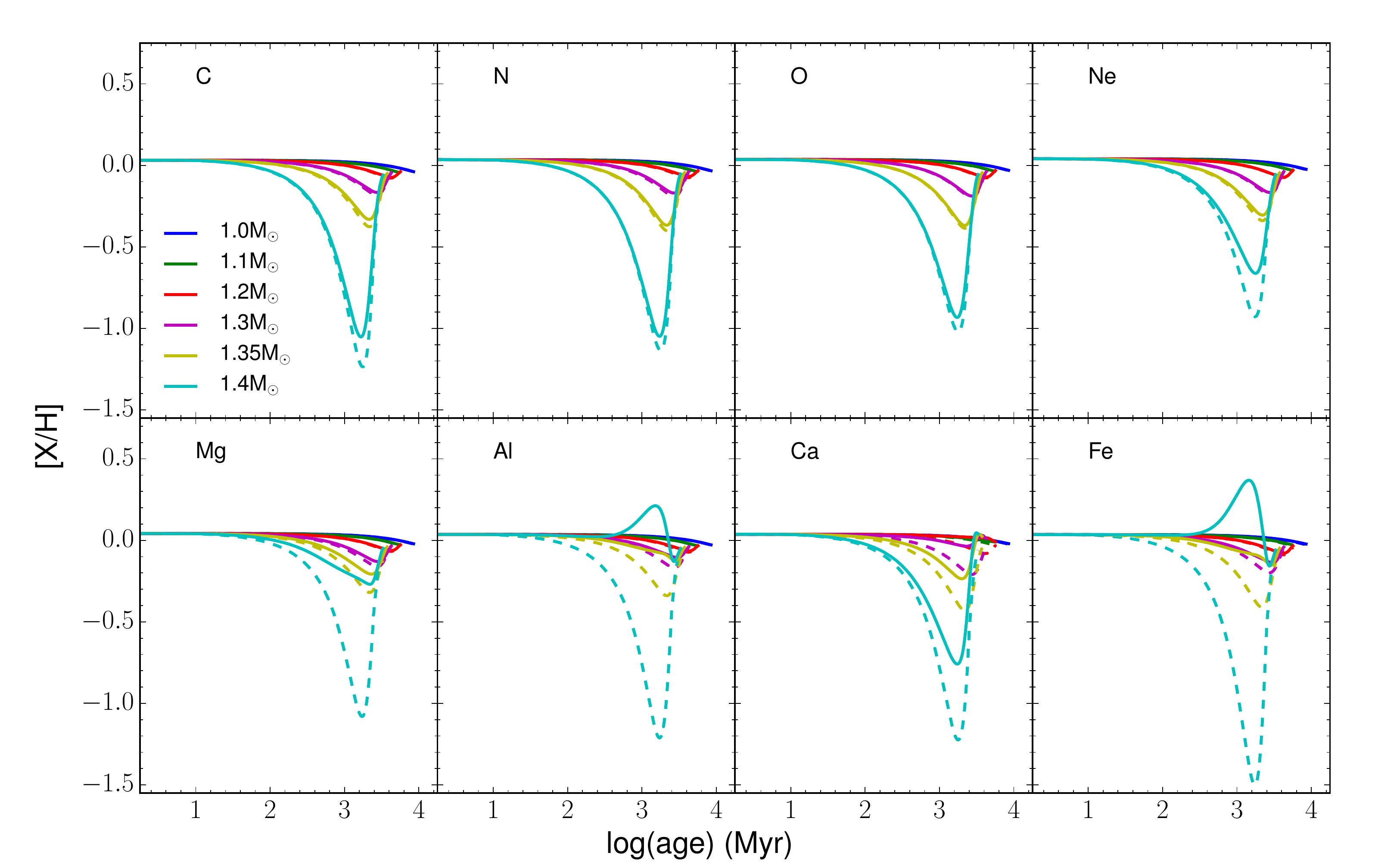}
\caption{Evolution of surface abundances ([X/H] calculated as in Eq.\ \ref{feh}) with time for eight elements for the grid 2 at solar metallicity. The solid and dashed curves respectively represent models with and without g$_\mathrm{rad}$. }
\label{fig:3}
\end{figure*}

Atomic diffusion processes are rather efficient in the outer layers of stars because diffusion time-scales are approximately proportional to the density of protons. For a given star, there  always exists a layer beyond which the diffusion time-scale is larger than the age of the considered star. If this limit layer is too close to (or above) the bottom of the outer convection zone, there is not enough time for atomic diffusion to play a significant role during the lifetime of the star. This is why the effects of atomic diffusion at solar metallicity are larger for stars with a solar mass \citep{turcotte98} or larger, i.e. for stars with a superficial convection zone that is not deeper than in the Sun. However, it should be noted that significant effects for lower mass stars cannot be excluded, since the age of these stars may be large enough \citep[see][]{dotter17}. Moreover, since at low metallicities, surface convective zones are shallower, atomic diffusion may therefore be efficient for lower masses \citep{richard02}. 

In Fig. \ref{fig:2}, the evolutionary tracks are shown for several initial metallicities (i.e. representative of the photosphere when abundances are still homogeneous outside the stellar core), and for masses ranging from 0.9 to 1.5~M$_{\odot}$. For the lowest metallicity ([Fe/H]$_{ini}=-0.35$), the role of g$_\mathrm{rad}$ appears well pronounced for masses larger than 1.1~M$_{\odot}$. This lower mass threshold is 1.3~M$_{\odot}$ for [Fe/H]$_{ini}=0.035$, and  1.45~M$_{\odot}$ for [Fe/H]$_{ini}=+0.25$. The role of g$_\mathrm{rad}$ is stronger at low metallicity because g$_\mathrm{rad}$ are larger for smaller abundances. This is a radiation transfer effect, since the momentum transfer between the net radiation flux and the considered element is strongly dependent on the saturation effect of bound-bound atomic transitions \citep{Alecian00}.

\subsection{Abundance variations}\label{ab}

Competition between gravity and g$_\mathrm{rad}$ leads to a migration of the chemical elements inside stable zones (when no mixing is at work) of the stars. When g$_\mathrm{rad}$ are not taken into account, all the elements (except hydrogen) migrate toward the center of the star due to gravitational settling, and this may cause strong depletion of metals at the surface. Therefore, taking into account g$_\mathrm{rad}$ generally prevents this abnormal superficial depletion (see Ne, Mg and Ca in Fig. \ref{fig:3}). In some cases g$_\mathrm{rad}$ are so large at the bottom of the surface convection zone, that metals enter the convection zone and their superficial abundances increase (see Al and Fe in Fig. \ref{fig:3} for the 1.4~M$_{\odot}$ case).

\begin{figure*}
\center
\includegraphics[scale=0.55]{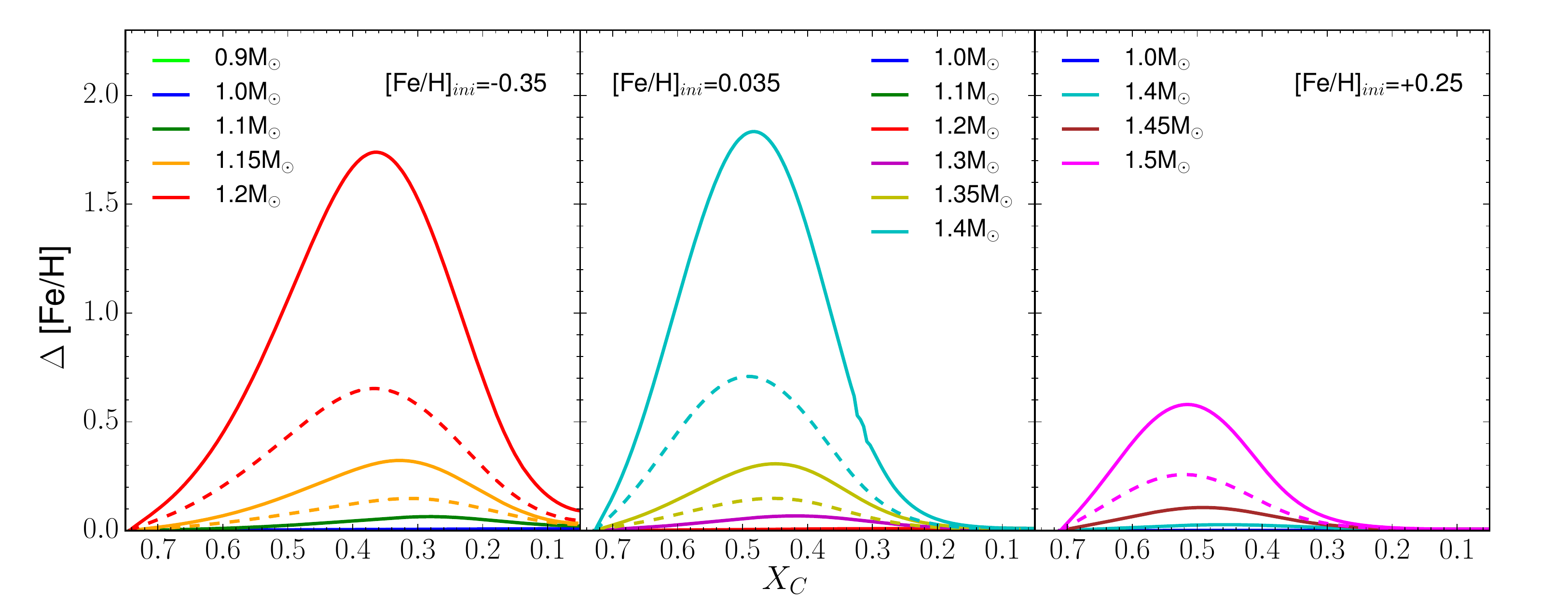}
\caption{Evolution of the difference in [Fe/H] between models with and without g$_\mathrm{rad}$ for the three grids of models. $X_C$ is the central hydrogen mass fraction. The solid lines show the differences for models including the effect of partial ionisation while the dashed lines show the differences when this process is not taken into account.}
\label{fig:4}
\end{figure*}

These changes of element distribution inside the star, iron in particular, explain the slightly different evolution of the models in Fig. \ref{fig:2}. This shows that [Fe/H], an observable parameter characterizing stars, may be affected by the inclusion of g$_\mathrm{rad}$. When [Fe/H] is used as an observational constraint in stellar evolution calculation to determine unknown stellar parameters like age or mass, the error in that determination will likely be larger if the grid is computed without the effect of g$_\mathrm{rad}$ (see Section \ref{opti}). In our three grids, the difference in [Fe/H] goes from 0 to 1.7 dex (see Fig. \ref{fig:4}) between the models with and without g$_\mathrm{rad}$. As discussed previously, the effect of g$_\mathrm{rad}$ for the largest metallicity grid (grid 3) is lower than for the others and it is visible in the difference in [Fe/H]. Despite this, the difference in [Fe/H] is larger for the model with 1.4~M$_{\odot}$ of grid 2 than for the model with 1.2~M$_{\odot}$ of grid 1 even if g$_\mathrm{rad}$ are more efficient for models of grid 1. This is due to the deepening of the surface convection zone which is larger at low metallicity and dilute the accumulated iron more efficiently in the surface convective zone (see Section \ref{ZC}).

The surface abundances of some elements (He, C, N and O for instance) in our computations are not representative of the ones obtained from the observations of G and F type stars (at least during a fraction of the evolution of the models). The maximum depletion observed for these elements is $\approx 0.4$ dex for star with a solar metallicity \citep[see][]{adibekyan12,bensby14,brewer16}. These elements are not or only weakly supported by radiative accelerations and are largely depleted in the models even when g$_\mathrm{rad}$ are taken into account. This result is expected because these models do not include additional mixing processes (mixing induced by rotation for example) which should reduce these large depletions. The abundances of the present study may then be considered as upper limits of what can be obtained from more complete models including atomic diffusion and competing macroscopic processes.

\subsection{Position of the bottom of the surface convection zone}\label{ZC}

In the mass range covered by our model sample, the main abundance differences between the two sets of models occur inside the convection zones due to the diffusion flux of iron at their bottom. There is no significant accumulation of metals in layers below the surface convection zone where atomic diffusion processes are too slow to produce abundance stratifications, contrarily to what happens in A and B type stars \citep{richard01,theado09,deal16}. Here the structure of the models is modified only near the stellar surface.

\begin{figure}
\center
\includegraphics[scale=0.47]{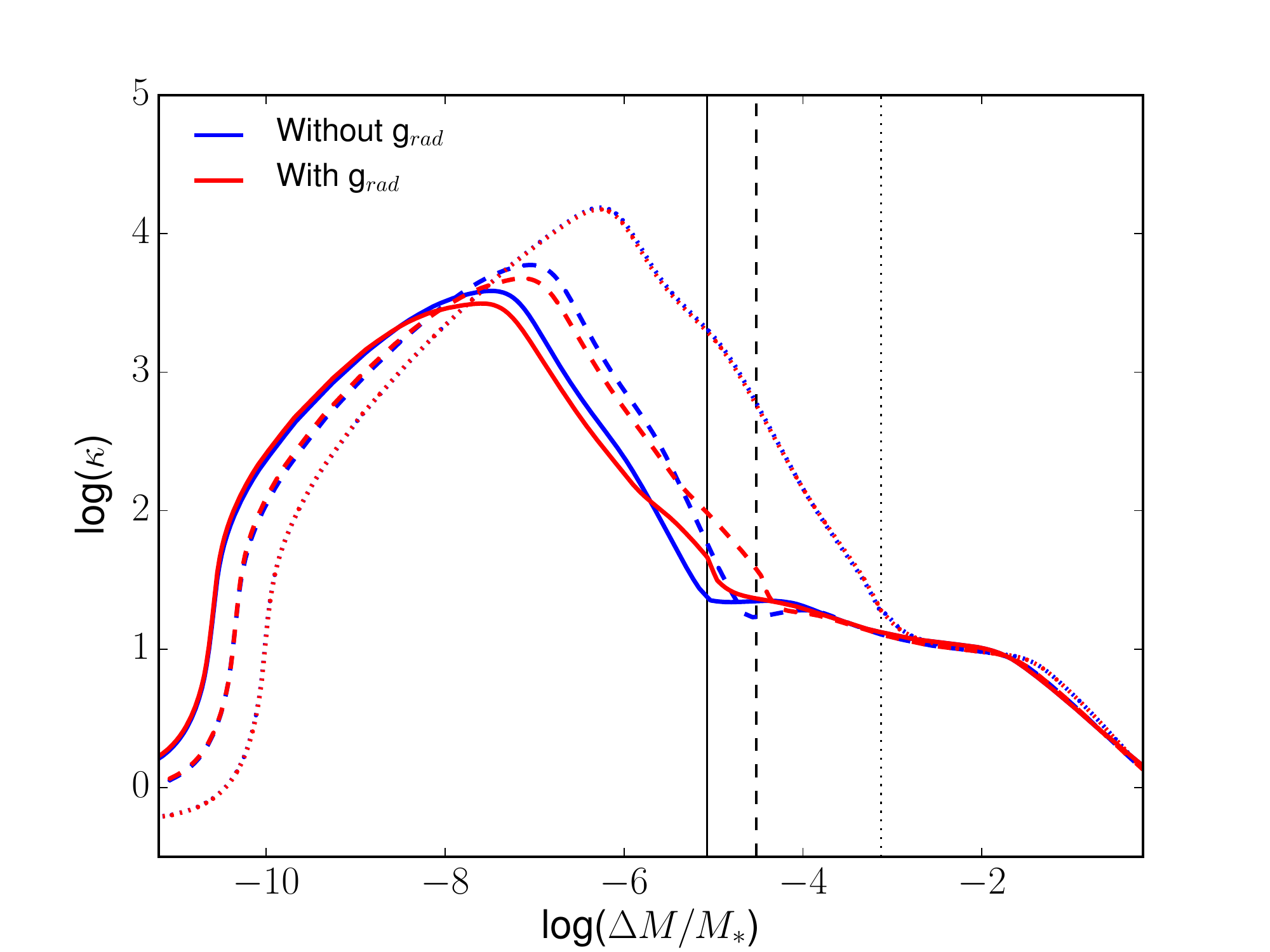}
\caption{Rosseland opacity profiles of the 1.4~M$_{\odot}$ of grid 2 for $X_C=0.6$ (solid curves), $X_C=0.4$ (dashed curves) and for $X_C=0.2$ (dotted curves). The blue and red curves represent respectively the models without and with g$_\mathrm{rad}$. The solid dashed and dotted vertical lines represent the position of the bottom of the surface convection zone for the model without g$_\mathrm{rad}$ for the same value of $X_C$ than the opacity profiles (they are not represented for the model with g$_\mathrm{rad}$ for clarity).}
\label{fig:5}
\end{figure}

The accumulation of iron, aluminium (model 1.4~M$_{\odot}$ of grid 2, see Fig. \ref{fig:3} for example) and calcium (model 1.2~M$_{\odot}$ of grid 2, see Fig. \ref{fig:3} for example), or the depletion for the other elements has a direct influence on the Rosseland opacity. Figure \ref{fig:5} shows the Rosseland mean opacity profiles of 1.4~M$_{\odot}$ models with and without g$_\mathrm{rad}$. The difference is more important close to the bottom of the surface convection zone (increase of 65\% at ${X_c=0.4}$) and this has a direct influence on the evolution of the star (i.e. for the structure) and the surface abundances. As iron is one of the main contributors to the opacity in this region, its accumulation leads to a larger opacity in this region compared to that obtained with gravitational settling alone.

\begin{figure*}
\center
\includegraphics[scale=0.55]{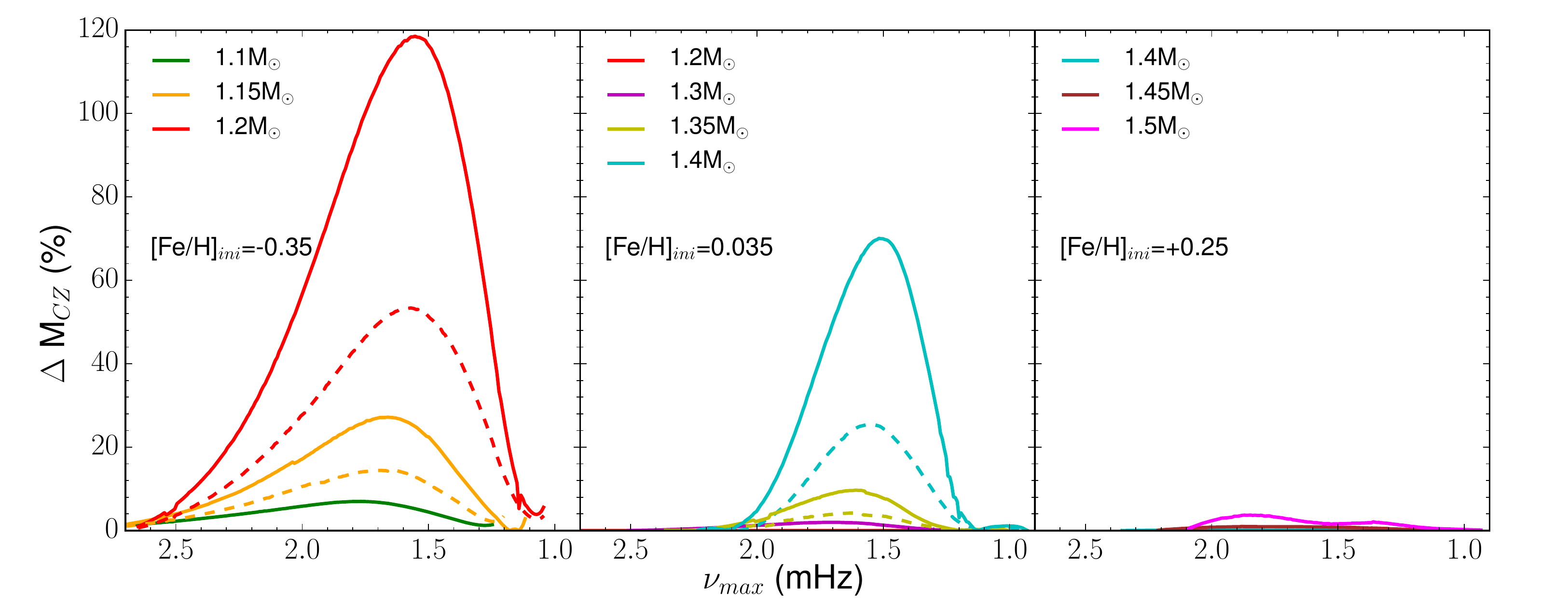}
\includegraphics[scale=0.55]{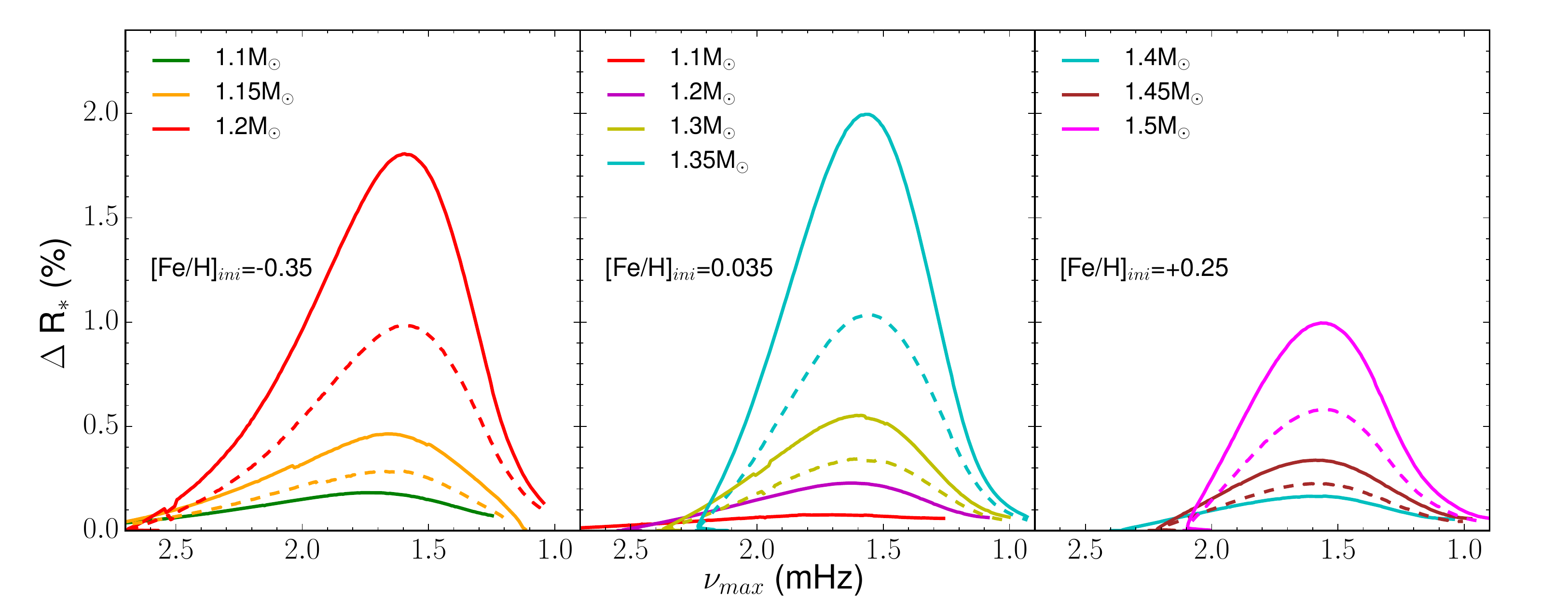}
\caption{Evolution of the difference of the mass of the surface convection zone (upper panels) and of the difference of the radius of the models (lower panels) with the frequency at maximum power $\nu_{max}$ (see Section \ref{sismo0}) for the three grids of models. Dashed lines are for the same models but without the effect of partial ionisation.}
\label{fig:6}
\end{figure*}

As a result, the bottom of the surface convection zone is always deeper when g$_\mathrm{rad}$ are taken into account (see upper panels of Fig. \ref{fig:6}) as it was already shown by \cite{turcotte98b} for F type stars. The more massive the star, the more important is the deepening of the surface convection zone due to g$_\mathrm{rad}$. Once again this effect is larger for lower metallicity stars. This maximum difference which can be obtained from models with and without g$_\mathrm{rad}$ reaches 120\% for grid 1 and goes down to 65\% and 5\% for grids 2 and 3 for the more massive models of the three grids.

Note that the deepening of the convection zone is smaller in our models than in \citet{turcotte98b} models. We presume that this could be due to the fact that the radiative acceleration for Ni, which significantly contributes to the opacity, is presently missing in our calculations. The new SVP tables that are in preparation (private communication of Alecian and LeBlanc, 2018), should improve our models in the near future.

\subsection{Variation of the stellar radius}

We have seen in previous sections that the accumulation of metals modifies superficial abundances, opacity profiles and size of the convection zone. Since the structure of the star is modified, so is the radius. Accurate knowledge of the radius is important to characterize exoplanets found by the transit method. If we compare the stellar radii computed without g$_\mathrm{rad}$ to those computed with g$_\mathrm{rad}$ (see lower panels of Fig. \ref{fig:6}), models with g$_\mathrm{rad}$ always give larger radii. The maximum difference which can be obtained from models with and without g$_\mathrm{rad}$ never exceeds 2\% and is at the level of requested uncertainties for the PLATO objectives. The increase of radius in our g$_\mathrm{rad}$ models is linked to a decrease of the mean density due to atomic diffusion including g$_\mathrm{rad}$, the same effect however smaller in magnitude, was found for the Sun by \cite{turcotte98}. 

\section{Seismic implications}\label{sismo0}

\begin{figure*}
\center
\includegraphics[scale=0.55]{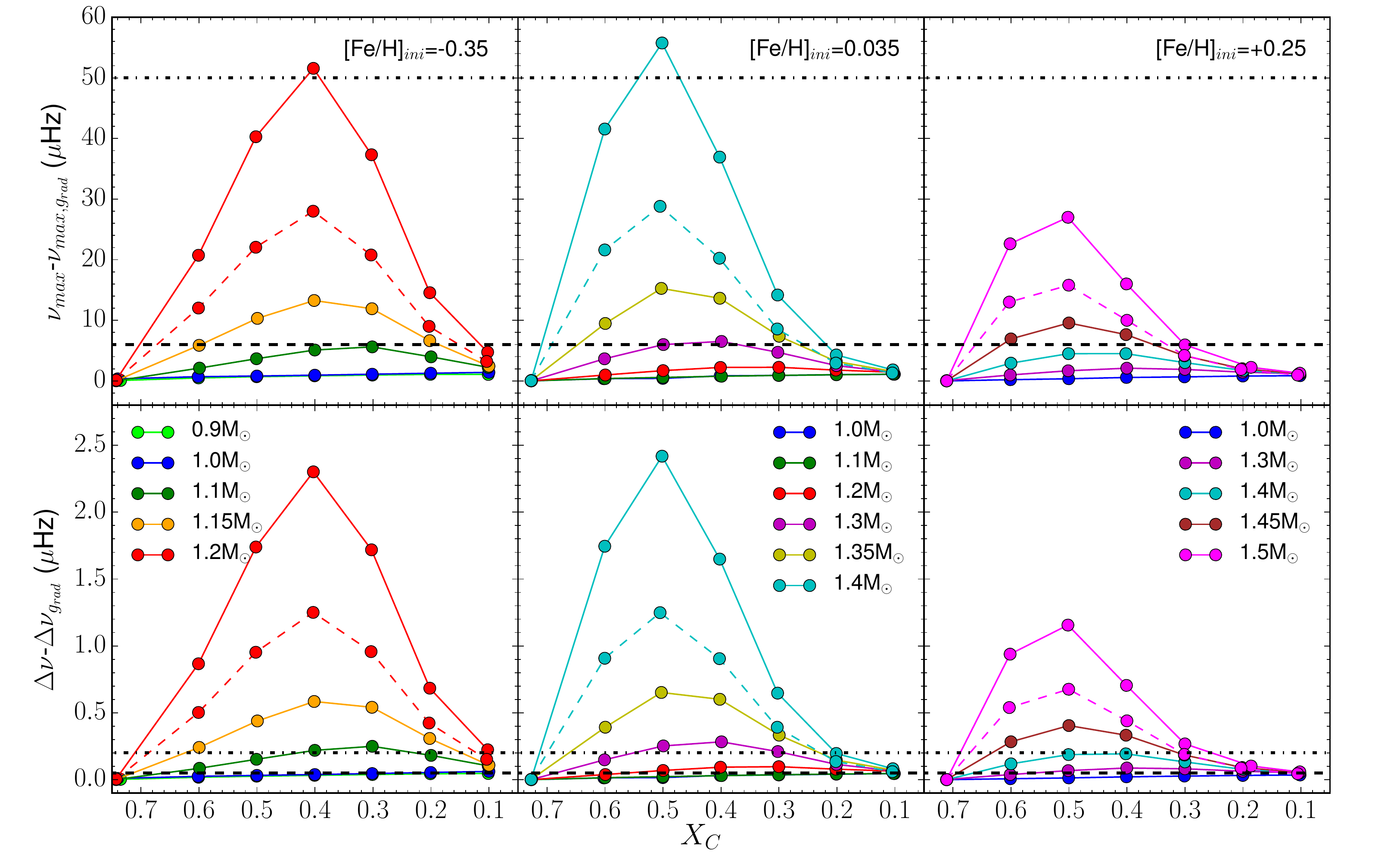}
\caption{Evolution with the central hydrogen content of the differences of frequency at maximum power, $\nu_{max}$, between models without and with g$_\mathrm{rad}$ for the three grids (upper panels). Same for the average large separation $\Delta \nu_0$ (lower panels). Each colour corresponds to a given mass. The dashed lines represent the same models but without the effect of partial ionisation. The horizontal black dashed-dotted lines indicate the adopted A uncertainty set and the horizontal black dashed lines indicate the adopted B uncertainty set on $\nu_{max}$ (upper panels) and $\Delta \nu_0$ (lower panels).}
\label{fig:7}
\end{figure*}

Our study confirms that g$_\mathrm{rad}$ may have non negligible effects on stars, especially on the iron surface abundance and on the size of the surface convection zone. Can these changes have detectable effects on the seismic properties of the star? We consider here only the global seismic indices, leaving a more-in depth study of individual frequencies and frequency combinations for a forthcoming paper. 
The global asteroseismic indices are the frequency at maximum power, $\nu_{max}$ and the averaged large frequency separation $\Delta \nu_0$
\citep{chaplin13}. Scaling relations relating these seismic indices to stellar mass, radius and effective temperature are expressed for solar-like oscillating main-sequence stars as \citep{kjeldsen95}:

\begin{equation}
\nu_{max}=\left(\frac{M}{M_\odot}\right)\left(\frac{R}{R_\odot}\right)^{-2}\left(\frac{T_{\mathrm{eff}}}{5777~K}\right)^{-\frac{1}{2}}~3.05~\text{mHz}
\end{equation}

\begin{equation}
\Delta \nu_0=\left(\frac{M}{M_\odot}\right)^{\frac{1}{2}}\left(\frac{R}{R_\odot}\right)^{-\frac{3}{2}}~134.9~\mu\text{Hz}
\end{equation}

We showed that g$_\mathrm{rad}$ have an impact on $T_\mathrm{eff}$ and on the radii of stars for a given mass (Section \ref{PLATO}), so an effect should be visible in the $\nu_{max}$ and $\Delta \nu$ values. In order to be detectable, the seismic signatures of the g$_\mathrm{rad}$ must be larger than the uncertainties arising from the observations. The  \textit{Kepler Legacy} sample of solar-like oscillating stars includes stars in the mass range $0.8 -1.6$~M$_\odot$ with [Fe/H] in the range [-1,+0.5] dex \citep{silva17}. For most of these stars, \cite{lund17} obtained uncertainties for $\nu_{max}$ and $\Delta \nu$ in the approximate range 6-50 $\mu$Hz and 0.05-0.2 $\mu$Hz respectively, depending on the apparent magnitude (in the range 6-11 mag) and the observing time (between 12 months and more than four years). The PLATO mission aims to measure individual frequencies of a reference star (1~M$_\odot$, 1~R$_\odot$, 6000K) with uncertainties no larger than $0.2~\mu\text{Hz}$ at magnitude 10 \citep{rauer14}. The PLATO uncertainties for $\nu_{max}$ and $\Delta \nu$ are expected to lie in similar ranges than \textit{Kepler} at given magnitude but PLATO will observe a larger number of bright stars and therefore with expected uncertainties in the lower side of the range. For purpose of comparison, we therefore considered two sets of uncertainties on $\nu_{max}$ and $\Delta \nu$ (see Table \ref{table:3}). The first set (A) is based on the uncertainties of the best \textit{Kepler Legacy} data \citep{silva17,lund17} and the bulk of bright PLATO target stars. The second set (B) considers more conservative uncertainties.
We hereafter compare the effects of g$_\mathrm{rad}$ on $\nu_{max}$ and $\Delta\nu_0$ to the aforementioned uncertainties.

\subsection{g$_\mathrm{rad}$-induced change on $\nu_{max}$ and $\Delta\nu_0$}\label{sismo}
Figure \ref{fig:7} compares $\nu_{max}$ and $\Delta\nu_0$ for our selected sample of masses and metallicities at seven stages along the main-sequence. We find that the values of $\nu_{max}$ and $\Delta\nu_0$ are always smaller for models including g$_\mathrm{rad}$. 

\begin{table}
\centering
\caption{Considered uncertainties on observed $\nu_{max}$ and $\Delta\nu_0$}
\label{table:3}
\begin{tabular}{l c c  }
\hline
Uncertainty sets (in $\mu$Hz) & $\delta \nu_{max}$ & $\delta \Delta\nu_0$  \\
\hline
\hline
A & 6 & 0.05 \\
B & 50 & 0.2 \\     
\hline
\end{tabular}
\end{table} 

\begin{table}
\centering
\caption{Mass above which g$_\mathrm{rad}$ have non-negligible effect on seismic predictions} 
\label{table:4}
\begin{tabular}{l c c c }
\hline
Grid & 1 & 2 & 3 \\
\hline
\hline
[Fe/H]$_{ini}$ & -0.35 & 0.035 & +0.25 \\
Limit Mass A\tablefootmark{a} (M$_{\odot}$) & 0.9 & 1.1 & 1.2 \\
Limit Mass B\tablefootmark{b} (M$_{\odot}$) & 1.05 & 1.25 & 1.4 \\
     
\hline
\end{tabular}
\tablefoot{\tablefoottext{a}{Limit determined from the differences obtained in $\Delta\nu_0$ with an uncertainty of $0.05 \mu\text{Hz}$ (A).}\tablefoottext{b}{Limit determined from the differences in $\Delta\nu_0$ with an uncertainty of $0.2 \mu\text{Hz}$ (B).}
}
\end{table} 

Regarding $\nu_{max}$, the impact of g$_\mathrm{rad}$ never exceeds $15~\mu\text{Hz}$ except for the most massive models. This is more than 3 times lower than the B set of uncertainties but 2.5 times larger than the A set of uncertainties. We conclude that g$_\mathrm{rad}$ need to be very efficient in order to produce a significant signature in the $\nu_{max}$ value.

The effects on $\Delta\nu_0$ are more important. The inclusion of g$_\mathrm{rad}$ leads to differences that reach up to $2.4~\mu\text{Hz}$ (for the model of 1.4~M$_{\odot}$ at solar metallicity), that-is much larger than any uncertainty derived from Kepler data or expected from PLATO data. In fact, 
$\Delta\nu_0$ being directly related to the mean density of the star, differences in radius as small as 2\%  can still induce large differences in $\Delta\nu_0$.

We may now define the limit mass $M_L$ as the stellar mass above which the change of $\Delta\nu_0$ due to g$_\mathrm{rad}$ is larger than the uncertainty sets A and B. Considering the A set, the values of $M_L$ are 1.05, 1.25, and 1.4 $M_\odot$, for grid 1, 2, and 3 respectively. In the case of the B set, $M_L$ are lower (0.9, 1.1, and 1.2 $M_\odot$ for grid 1, 2, and 3, respectively). These values of $M_L$ are listed in Table \ref{table:4}, and will serve as references. They correspond to the smallest masses below which g$_\mathrm{rad}$ can be neglected. For masses larger than these limits, the effect of g$_\mathrm{rad}$ will depend on the efficiency of other transport processes.

\subsection{g$_\mathrm{rad}$-induced uncertainties on seismic ages}

When modelling a star using seismic constraints, the impact of g$_\mathrm{rad}$ on $\nu_{max}$ and $\Delta\nu_0$ generates an uncertainty on the age of the star. An order of magnitude of the age uncertainty can be 
obtained for instance by comparing the ages of standard and g$_\mathrm{rad}$ models at fixed mass, metallicity, central hydrogen abundance  and $\Delta\nu_0$. 
In such a configuration, we find that the age of the model with g$_\mathrm{rad}$ is always smaller than that of the standard model in this study.

The maximum difference due to g$_\mathrm{rad}$ at metallicity [Fe/H]$_{ini} = 0.035$ (grid 2) is obtained for the 1.4~M$_\odot$ model at $X_C=0.4$ and $\Delta\nu_0=82.90~\mu\text{Hz}$.
The ages of the corresponding standard and g$_\mathrm{rad}$ models are respectively  $1.546~\text{Gyr}$ and $1.386~\text{Gyr}$, that-is they differ in age by about 9\%. Similarly for the most massive models of grid 1 and 3, we obtain age differences of about 6\% and 5\%. The g$_\mathrm{rad}$ therefore contribute for a significant part to the age error budget for most massive main-sequence stars showing solar-like oscillations. 

\subsection{Acoustic depths of the base of the convection zone}

In Section \ref{ZC}, we showed that the depth of the surface convection zone increases when g$_\mathrm{rad}$ are included. The question then is whether the g$_\mathrm{rad}$-induced change of the size of the CZ is significant.
Solar-like oscillations enable the  measurement of the acoustic depth of the base of the convection zone which is defined as : 
\begin{equation}
\tau_{CZ,obs} =\int^{R_\ast}_{r_{CZ}}~\mathrm{dr}/c_s
\end{equation}\noindent where $r_{CZ}$ is the radius of the bottom of the surface CZ, $c_s$ the sound speed and $R_\ast$ the radius of the star \citep[][and references therein]{mazumdar01}. 
We therefore computed the acoustic depths, $\tau_{CZ,obs}$, for our models and compared the resulting g$_\mathrm{rad}$-induced differences $\Delta \tau_{CZ,RA}$ to the observational uncertainties of seismically measured $ \tau_{CZ,RA}$. 
From our models, we find that the maximum g$_\mathrm{rad}$-induced differences for the convective sizes roughly correspond to $\Delta \tau_{CZ,RA} \sim 300$~s  for the 1.4~M$_{\odot}$ model of grid 2 and to 340~s for the 1.2~M$_{\odot}$ model of grid 1 at fixed $X_C$. This difference goes down to 160~s for the first case when comparing models with the same radius. Seismically measured $\tau_{CZ,obs}$ were obtained by \cite{verma17} for stars from the \textit{Kepler Legacy} sample. These authors found typical uncertainties on  $\tau_{CZ,obs}$ of the order of 150~s for stars with masses of about 1.4~M$_{\odot}$ and of the order of 75~s for stars with masses of about 1.2~M$_{\odot}$. Thus, we can conclude that g$_\mathrm{rad}$ must be taken into account in the models for determining the properties at the base of the convection zone for the most massive stars in the range of interest showing solar-like oscillations.  

\section{Impact of g$_\mathrm{rad}$ on [Fe/H] and on the stellar parameter determinations}\label{opti}

\begin{figure}
\center
\includegraphics[scale=0.47]{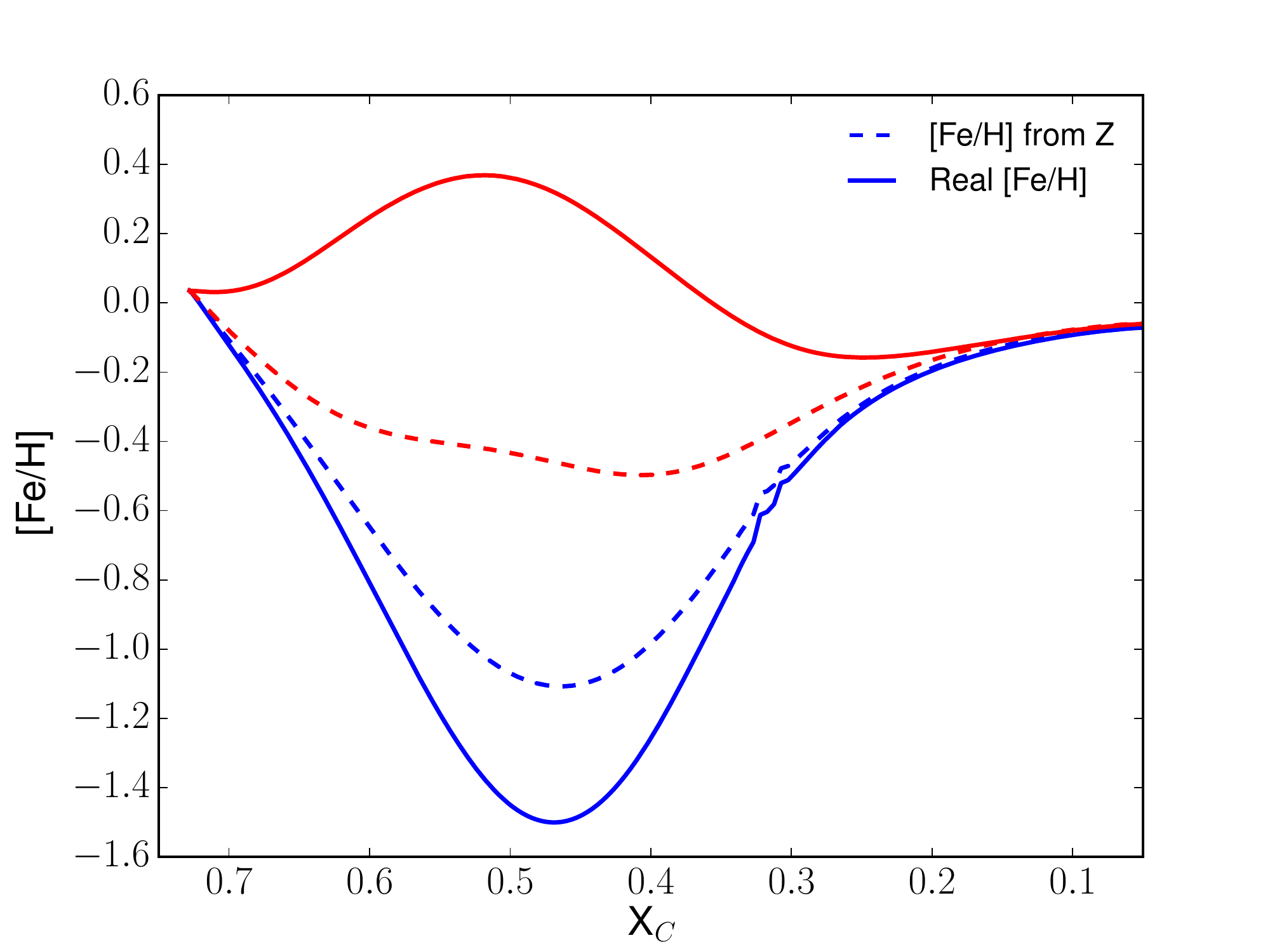}
\caption{Comparison between [Fe/H] computed with the real surface abundance of iron and hydrogen (solid lines) and computed from the metallicity value (dashed lines) for 1.4~M$_{\odot}$ models with (red curves) and without (blue curves) g$_\mathrm{rad}$. }
\label{fig:8}
\end{figure}

With CoRoT and Kepler high quality seismic data, it is possible to determine very precise stellar parameters such as masses, radii and ages  
for solar-like oscillating dwarfs \citep{lebreton14, silva17, reese16}. In that framework, one significant impact of the g$_\mathrm{rad}$ on the stellar parameter determination is its effect on the relation between the iron content and the metallicity.

A stellar parameter determination nowadays is usually achieved by means of an optimisation process. This method looks for the stellar model that best fits the observed oscillation frequencies and/or frequency combinations and additional spectroscopic constraints such as the effective temperature and/or $\log ~ g$. 
The stellar model computations involved in the best-fit search require the knowledge of the initial metallicity $Z_{ini}$.
However the available spectroscopic constraint which is used for the best-fit search is the surface iron abundance of the star 
[Fe/H] determined from observations. Assuming a chemical mixture scaling, one derives the current surface metallicity $Z_s$. However this quantity can significantly differ from the initial metallicity $Z_{ini} $  of the star due to internal transport processes occurring over time. In particular, g$_\mathrm{rad}$ can lead to an accumulation of iron at the surface. This means that we must expect a smaller initial iron abundance than the observed one. When only gravitational settling is taken into account, the effect is the opposite.

In addition to these difficulties, we emphasize that atomic diffusion, especially g$_\mathrm{rad}$, acts differently on the chemical elements. Then when iron  accumulates at the surface of the star, it is no longer possible to approximate the surface metallicity  $Z_s$ using the determination of [Fe/H] by spectroscopy. Figure \ref{fig:8} compares the values of [Fe/H] considering: 
\begin{itemize}
\item the surface abundances of iron and hydrogen following Equation \ref{feh},
\item $\mathrm{[Fe/H]=[M/H]}$, i.e. $\mathrm{[Fe/H]}$ is assimilated to the surface metal to hydrogen abundance ratio.
\end{itemize}

When considering only gravitational settling (blue curves), the difference between the two computation methods gives similar evolutions of the profiles for a 1.4~M$_{\odot}$. Nevertheless there are differences up to 0.4 dex that-is much larger than current observational uncertainties. As the elements are diffusing toward the center but at different velocities, the scaling of the iron abundance with $Z$ is not possible even in that case. The difference reaches 0.7~dex for the models including g$_\mathrm{rad}$ (red curves) and the evolution is completely different as the iron is accumulated at the surface. In this case iron does not follow the behaviour of other heavy elements (namely CNO) for which gravitational settling is dominating the diffusion. It is clear in this example that the [Fe/H] value needs to be computed with the actual value of iron and hydrogen abundances. The differences between the two methods to compute [Fe/H] are smaller for lower mass stars and/or when other transport processes are taken into account since atomic diffusion is less effective. This issue needs to be investigated especially in the framework of optimisation methods as evolution codes used to compute stellar models rarely follow the evolution of the iron abundance.

\section{Discussion}

\subsection{Impact of partial ionisation}
In all the comparisons we have made on the structural and seismic properties, we observe that neglecting partial ionisation strongly underestimates the impact of atomic diffusion, especially for the most massive stars of our grids. As shown in Fig. \ref{fig:4}, \ref{fig:6} and \ref{fig:7}, the impact is roughly doubled when partial ionisation is taken into account. This is because iron dominates the structure modifications, and because it is among the elements we consider, the one for which neglecting partial ionisation in estimating the mean electric charge induces the largest errors (it has the highest atomic number). It is clear from this study that partial ionisation must be taken into account in modelling main-sequence stars.

\subsection{Effect of the initial solar mixture}
We demonstrated how the initial metallicity is an important parameter in evolution models including g$_\mathrm{rad}$. To evaluate the impact of the adopted solar mixture, we compared models based on the solar mixture of AGSS09 to models based on \cite{grevesse93} (hereafter GN93 mixture). We computed two 1.3~M$_{\odot}$ models with the GN93 mixture, with and without g$_\mathrm{rad}$ in order to perform the same comparisons as in Section \ref{sismo}. In these two models $(Z/X)_{ini}=0.0276$ and $\alpha_{CGM}=0.678$ as inferred from a solar calibration.

The solar metallicity of the GN93 mixture is larger than the AGSS09 one. We showed in previous sections that g$_\mathrm{rad}$ decrease when the metallicity increases for a given mass. Therefore, the effects of g$_\mathrm{rad}$ are slightly smaller in models using the GN93 mixture but still remain non-negligible. With the GN93 mixture, the mass above which g$_\mathrm{rad}$ have non-negligible effects on seismic predictions is only $\approx 0.05$~M$_{\odot}$ higher than the limit mass obtained with the AGSS09 mixture (Table \ref{table:4}). The difference for other solar mixtures \citep{grevesse98,asplund05} is expected to be smaller as the metallicity difference with AGSS09 is smaller.
 
\subsection{Implication for the PLATO space mission}

\begin{figure}
\center
\includegraphics[scale=0.47]{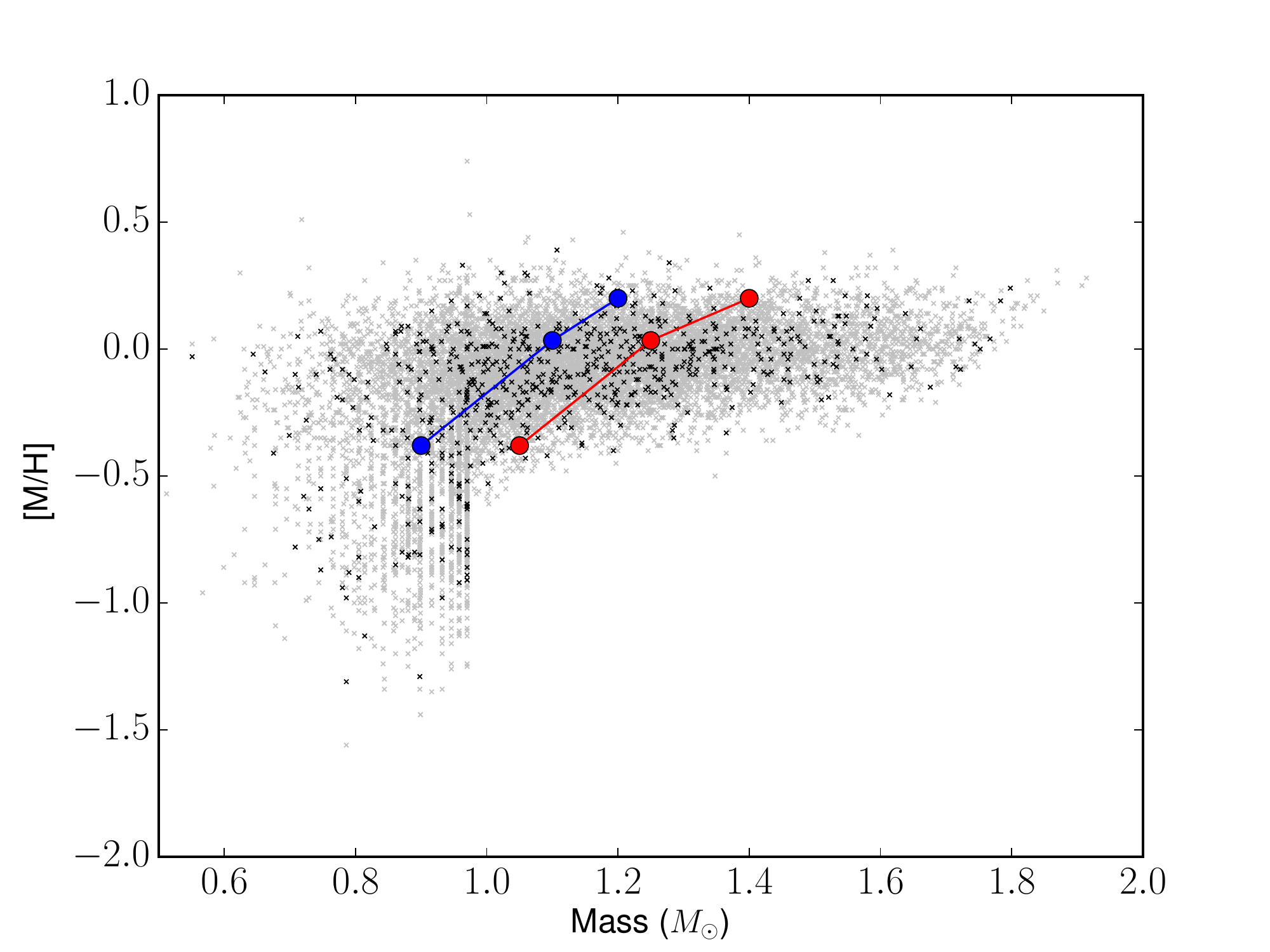}
\caption{Metallicity according to the mass of a population simulation of the PLATO (grey crosses) and \textit{Kepler} (black crosses) core programme stars. The selected stars are from K7 to F5 with magnitude between 4 < V < 11, effective temperature between 4030 < $T_{\mathrm{eff}}$ < 6650 K, and luminosity classes between IV and V. The blue and red points correspond to the models listed Table \ref{table:4} which represent masses when g$_\mathrm{rad}$ needs to be taken into account.}
\label{fig:9}
\end{figure}

In Section \ref{sismo}, we determined that g$_\mathrm{rad}$ induce differences in
$\nu_{max}$ and $\Delta\nu_0$ that can be larger than their observational uncertainties when the stellar mass lies above a lower limit mass $M_L$, which depends on the metallicity (Table \ref{table:4}). These lower masses can be used to determine whether g$_\mathrm{rad}$ have to be taken into account or not to ensure a given accuracy on the inferred stellar parameters. 
We may estimate the number of stars of the PLATO core program which might be affected by g$_\mathrm{rad}$. For this purpose, we use a stellar population  synthesis computed with the Besan\c{c}on Galaxy model \citep{robin03, czekaj14,robin14} (A. Robin, private communication). 
The simulation is representative of one PLATO observation field. The limit masses of Table \ref{table:4} are indicated by yellow (B set of uncertainties) and orange (A set of uncertainties) points in Fig. \ref{fig:9}. The number of stars with masses larger than the limit masses amounts to 33\% up to 59\% (depending of the uncertainty criteria) of the PLATO core program star sample and reaches 58\% up to 75\% for the total field. This number is an upper limit, but nevertheless indicates that for a significant number of stars, g$_\mathrm{rad}$ may not be negligible and the determination of their parameters will require some care if one wants to achieve the requested PLATO accuracy. 

\section{Conclusion}
We improved the CESTAM code in order to compute models including the effects of radiative accelerations on the chemical element profiles and the resulting effects on opacities. 
 
The goal was to characterize the sole transport effect of atomic diffusion including radiative accelerations; therefore no macroscopic transport apart from convection was assumed. We computed two sets of models at three metallicities for masses ranging between 0.9 and 1.5~M$_{\odot}$. One set includes the effect of g$_\mathrm{rad}$ and the other set does not.  

The effects of radiative accelerations are larger at low metallicities and for the more massive stars considered here. The most obvious impact of radiative accelerations in stars is the modification of the surface abundances. For instance, this process is responsible for the surface abundances of chemically peculiar stars and we show here that it has also an impact for low mass oscillating main-sequence solar-like stars. The most important to follow is iron as it is one of the main contributors to opacity while the [Fe/H] value is an important input for the stellar modelling. We showed that when radiative accelerations on iron are not negligible, it is not correct to calculate the [Fe/H] of a model simply considering a scaling of the metal content. This is because the effect of radiative accelerations is selective, and even if iron accumulates at the surface, the surface metallicity decreases as most of the other elements are depleted. This may have an important impact on the stellar parameter determination as [Fe/H] is an observational input. The difference in [Fe/H] between models with and without radiative accelerations reaches 1.7~dex for the more massive models of the grids. 

We showed that the accumulation of elements in the surface convection zone (mainly iron) induces structure modifications. This is mainly due to the local increase of the opacity at the bottom of the surface convection zone as elements accumulate in regions where they are main contributors to the opacity. This local increase of the opacity leads to an increase of the size of the surface convection zone which can reach up to 120\% in mass. This represents an increase larger than 160~s when considering the position of the bottom of the surface convection zone in acoustic radius. This is larger than the uncertainties obtained for some F-type stars of the \textit{Kepler Legacy} sample and has to be further investigated. The modification of the radius of the star induced by the effects of radiative accelerations can reach 2\%.

Using scaling relations we showed that the frequency at maximum power $\nu_{max}$ of a model can be significantly affected by radiative accelerations for the more massive stars of our sample. Some models of our grid showed difference in the large frequency separation of pressure modes $\Delta\nu_0$ larger than the observational uncertainty. For masses larger than 0.9, 1.1 and 1.2~M$_{\odot}$ (considering uncertainties of the \textit{Kepler Legacy} sample) respectively for [Fe/H]$_{ini}=-0.35, +0.035$ and $+0.25$, radiative accelerations may have an impact on the age, mass and radius determinations exceeding the precision requested by the PLATO main objectives. These masses are slightly larger when considering more conservative uncertainties. This has consequences on the parameters to be determined from \textit{Kepler}, and future TESS and PLATO data. We estimated that radiative accelerations should not be negligible for 33\% up to 58\% (depending on the considered uncertainties) of the core program stars of \textit{Kepler} and PLATO.

One must stress that the radiative accelerations impact might be lowered when other processes are efficient in transporting material within stars such as mixing induced by rotation, turbulence, internal gravity waves to name a few. This is out of scope of this paper, but will be studied in a forthcoming paper. 
 
\begin{acknowledgements}

We acknowledge Annie Robin for providing us with the Besan\c{c}on Galaxy models and Thierry Morel for fruitful discussions.
This work was supported by CNES. 
We acknowledge financial support from "Programme National de
Physique Stellaire" (PNPS) of CNRS/INSU, France. 
This research was partially funded by the Natural Sciences and Engineering Research Council of Canada (NSERC). We thank Calcul Canada and Calcul Québec for computational resources. We acknowledge the referee for his careful reading and relevant suggestions which improved the paper.
\end{acknowledgements}

\bibliographystyle{aa} 
\bibliography{biblio.bib} 

\end{document}